\documentclass[12pt]{article}
\begin{document}
\begin{center}
{\bf \Large Refined Topological  Vertex\\
 and Duality of  Gauge Theories in Generic Omega Backgrounds}\\
 \vspace{3cm}
 {\large Kei Ito}\\
 \vspace{1cm}
 Omohi Faculty\\
 Nagoya Institute of Technology\\
 Gokiso-cho, Showa-ku\\
 Nagoya 466-8555,Japan\\
 email: ito.kei@nitech.ac.jp\\
 \end{center}
\vspace{1cm}
The partition functions of refined topological strings(A-models) are computed, which  
give rise to the circle-compactified five-dimensional supersymmetric linear quiver 
gauge theories in generic (not necessarily self-dual) Omega backgrounds. 
Based on the slicing independence conjecture of refined topological 
string partition functions, it is demonstrated explicitly that the duality exists between 
 $SU(N)^{M-1}$ and $SU(M)^{N-1}$ 
supersymmetric linear quiver gauge theories, even in generic Omega 
backgrounds.  It is found that the relations between string moduli and gauge 
moduli are deformed from the self-dual case.
However if the duality map which preserves the ratio of the Omega 
background parameters q and t, is considered, duality maps of the gauge 
moduli are not changed from the self-dual case.
\newpage
Some time ago, Bao et al.[1] showed that there exists a duality between five-dimensional
(5d) $\mathcal{N}=1,SU(N)^{M-1}$ and $SU(M)^{N-1}$ linear quiver gauge theories compactified 
on $S^1$ in the self-dual $\Omega$ background. The topological string(topological A-model) 
amplitudes for toric Calabi-Yau three-folds in self-dual $\Omega$ background can be computed by 
diagrams with ``ordinary" topological vertex. These amplitudes give rise to Nekrasov 
partition functions[2] of 5d, $\mathcal{N}=1$ linear quiver gauge theories compactified on 
$S^1$ in the self-dual $\Omega$ background, upon the M theory lift of the  ``geometric
engineering". By the reflection symmetry of the toric diagram and duality maps of the string
moduli, they derived the duality maps of the string moduli and demonstrated the gauge theory duality
in the self-dual $\Omega $ background .The purpose of the present article is to show that 
the duality exists even in the generic (not necessarily self-dual)$\Omega$ backgrounds.\\

In the case of the generic $\Omega$ backgrounds, the topological string
(topological A-model) amplitudes for toric Calabi-Yau three-folds can be computed 
by diagrams with ``refined"topological vertex.[3] It has slightly different properties.
Instead of being cyclically symmetric, one of its legs indicates a preferred direction.
However the full partition function should be invariant under a change of choice 
of the preferred direction, which is called ``slicing invariance conjecture" and has been
demonstrated to be true in many concrete examples.[3],[4]\\

With the refined topological vertex,
 we compute the  partition functions of ``refined" topological A-models, which give rise
 to 5d $\mathcal{N}=1$, $SU(N)^{M-1}$-type linear quiver gauge theories compactified on $S^1$ in 
 ``generic"(not necessarily self-dual) $\Omega$ backgrounds, upon the M-theory lift of
geometric engineering. Let us start with the computation of the sub-diagram 
  ``vertex of a strip geometry", following[5]
with ``refined" topological vertex[3]\\

\[C_{\lambda\mu\nu}(t,q)=(\frac{q}{t})^{\frac{||\mu||^2+||\nu||^2}{2}} t^{\frac{\kappa_\mu}{2}}P_{\nu^t}
(t^{-\rho}, q, t)\]
\[\times\sum_\eta (\frac{q}{t})^{\frac{|\eta|+|\mu|+|\nu|}{2}} s_{\lambda^t/\eta}(t^{-\rho}q^{-\nu}
)s_{\mu/\eta}(t^{-\nu^t}q^{-\rho}),\hspace{3cm}(1) \]
where the Macdonald function is given by,\\
\[P_{\nu^t}(t^{-\rho},q,t)=t^{\frac{||\nu||^2}{2}}\tilde{Z}_\nu(t,q)\]
with,\[\tilde{Z}_\nu(t,q)=\prod_{(i,j)\in \nu}(1-t^{\nu^t_j-i+1}q^{\nu_i-j})\]
Then the amplitude of the left strip geometry of figure 1 is,\\
 \[ H^{Y^{(i-1)}_1Y^{(i-1)}_2...Y^{(i-1)}_N}_{Y^{(i)}_1Y^{(i)}_2...Y^{(i)}_N} 
(t, q, Q^{(i-1)}_{m_1},Q^{(i-1)}_{F_1},Q^{(i-1)}_{m_2},Q^{(i-1)}_{F_2},...,Q^{(i-1)}_{F_{N-1}},Q^{(i-1)}_{m_N})\]
\[=\sum _{\mu_1,\nu_1,\mu_2,\nu_2,...,\nu_{N-1},\mu_N}
(-Q^{(i-1)}_{m_1})^{|\mu_1|}(-Q^{(i-1)}_{F_1})^{|\nu_1|}(-Q^{(i-1)}_{m_2})^{|\mu_2|}(-Q^{(i-1)}_{F_2})^{|\nu_2|}...\]
\[\times ......(-Q^{(i-1)}_{F_{N-1}})^{|\nu_{N-1|}}(-Q^{(i-1)}_{m_N})^{|\mu_N|}\]
\[\times C_{\mu^t_1\emptyset Y^{(i-1)}_1}(t,q)C_{\mu_1\nu^t_1 Y^{(i)t}}(q,t)
C_{\mu^t_2\emptyset Y^{(i-1)}_2}(t,q)C_{\mu_2\nu^t_2 Y^{(i-1)t}_2}(q,t)...\]
\[\times ......C_{\mu^t_N\nu_{N-1} Y^{(i-1)}_N}(t,q)C_{\mu_N\emptyset Y^{{(i)}t}}(q,t)\hspace{5cm}(2)\]
where, $\mu_\alpha$ and $\nu_\alpha$ are Young tableaux for the internal lines with corresponding 
K\"{a}hler parameter of the homology cycles being $Q^{(i-1)}_{m_\alpha}$and $Q^{(i-1)}_{F_\alpha}$
, respectively. We choose the preferred direction in the horizontal lines (figure 1). Substituting
the definition of the refined topological vertex, and computing the summation over products of
(skew-)Schur functions, we have,\\

\[H^{{Y^{(i-1)}_1}...{Y^{(i-1)}_N}}_{Y^{(i)}_1...{Y^{(i)}_N}}
=\frac{\prod ^N_{\alpha=1}q^{\frac{||Y^{(i-1)}_{\alpha}||^2}{2}}\tilde{Z}_{Y^{(i-1)}_{\alpha}}(t,q)
t^{\frac{{||Y^{(i)t}_{\alpha}}||^2}{2}}\tilde{Z}_{Y^{(i)t}_{\alpha}}(q,t)}
{\prod_{1\leq\alpha<\beta\leq N}[Y^{(i-1)}_{\alpha},Y^{(i-1)t}_{\beta}]_
{\sqrt{\frac{t}{q}}Q^{(i-1)}_{\alpha\beta}}
[Y^{(i)}_{\alpha},Y^{(i)t}_{\beta}]
_{\sqrt{\frac{q}{t}}Q^{(i-1)-1}_{m_{\alpha}}Q^{(i-1)}_{\alpha\beta}Q^{(i-1)}_{m_{\beta}}}}\]

\[\times \prod_{1\leq \alpha<\beta\leq N} [Y^{(i)}_{\alpha},Y^{(i-1)t}_{\beta}]_{Q^{(i-1)-1}
_{m_{\alpha}}Q^{(i-1)}_{\alpha\beta} }\prod_{1\leq\alpha\leq \beta\leq N} 
[Y^{(i-1)}_{\alpha},Y^{(i)t}_{\beta}]_{Q^{(i-1)}_{\alpha\beta}Q^{(i-1)}_{m_{\beta}}}\hspace{.5cm} (3)\]
where,\[Q^{(i)}_{\alpha\beta}=\prod^{\beta-1}_{a=\alpha} Q^{(i)}_{m_a}Q^{(i)}_{F_a}\]
and the symbol $[Y_{\alpha},Y_{\beta}]_Q$is defined by,
\[  [Y_{\alpha},Y_{\beta}] _Q =\prod^{\infty}_{i,j=1}(1-Qq^{i-\frac{1}{2}-Y_{\alpha,j}}t^{j-\frac{1}{2}-Y_{\beta, i}})\]

\newpage

\begin{picture}(300, 350)

\put(50,50){\line(1,1) {36}}
\put(86,136){\line(1,1) {36}}
\put(122,247){\line(1,1) {36}}
\put(161,86){\line(1,1) {36}}
\put(197,172){\line(1,1) {36}}
\put(233,283){\line(1,1) {36}}
\put(86,86){\line(0,1) {50}}
\put(197,122){\line(0,1) {50}}
\put(50,25){\line(0,1) {25}}
\put(122,172){\line(0,1) {25}}
\put(122,222){\line(0,1) {25}}
\put(158,283){\line(0,1) {25}}
\put(161,61){\line(0,1) {25}}
\put(233,208){\line(0,1) {25}}
\put(233,258){\line(0,1) {25}}
\put(269,319){\line(0,1) {25}}

\put(0,50){$...Y^{(i-1)}_1$}
\put(36,136){$...Y^{(i-1)}_2$}
\put(72,247){$...Y^{(i-1)}_N$}

\put(70,64){$Q^{(i-1)}_{m_1}$}
\put(106,150){$Q^{(i-1)}_{m_2}$}
\put(144,261){$Q^{(i-1)}_{m_N}$}
\put(56,111){$Q^{(i-1)}_{F_1}$}
\put(90,182){$Q^{(i-1)}_{F_2}$}
\put(90,227){$Q^{(i-1)}_{F_{N-1}}$}

\put(181,100){$Q^{(i)}_{m_1}$}
\put(217,186){$Q^{(i)}_{m_2}$}
\put(253,297){$Q^{(i)}_{m_N}$}
\put(167,147){$Q^{(i)}_{F_1}$}
\put(203,218){$Q^{(i)}_{F_2}$}
\put(201,263){$Q^{(i)}_{F_{N-1}}$}

\put(116,86){$Y^{(i)}_1$}
\put(152,172){$Y^{(i)}_2$}
\put(188,283){$Y^{(i)}_N$}

\put(116,106){$Q^{(i)}_{B_1}$}
\put(152,192){$Q^{(i)}_{B_2}$}
\put(188,303){$Q^{(i)}_{B_N}$}

\put(227,122){$Y^{(i+1)}_1...$}
\put(263,208){$Y^{(i+1)}_2...$}
\put(299,319){$Y^{(i+1)}_N...$}

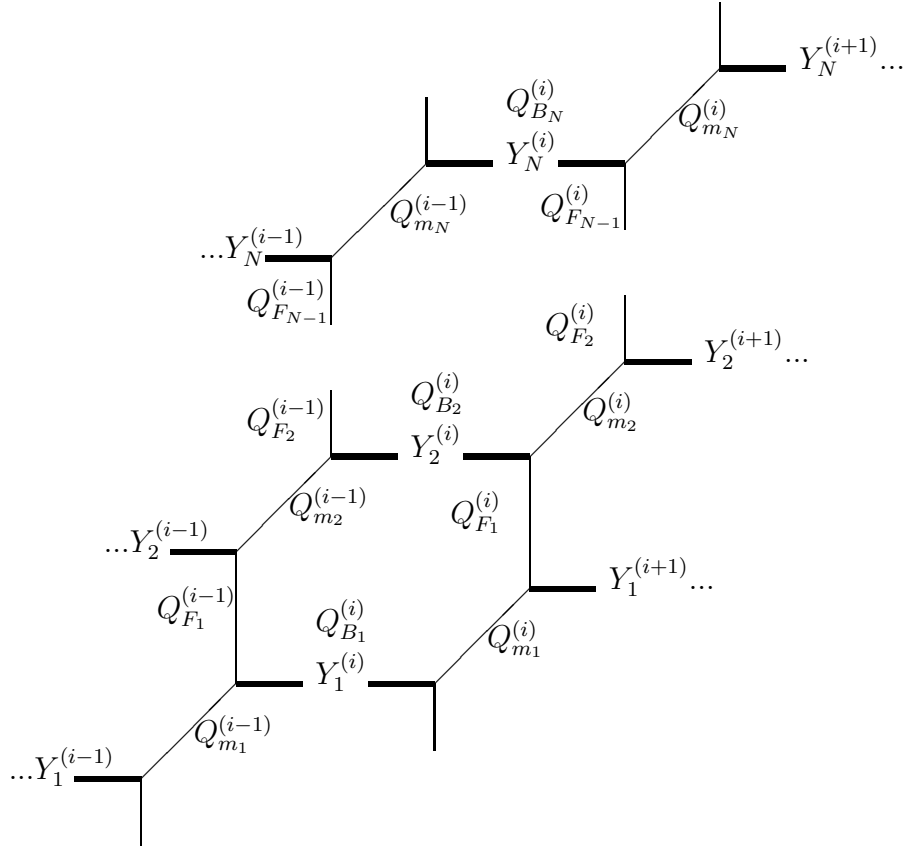
\put(0,10){Figure 1: The toric diagram which gives rise to $SU(N)^{M-1}$ theory}

\linethickness{2pt}
\put(25,50){\line(1,0) {25}}
\put(86,86){\line(1,0) {25}}
\put(61,136){\line(1,0) {25}}
\put(122,172){\line(1,0) {25}}
\put(97,247){\line(1,0) {25}}
\put(158,283){\line(1,0) {25}}
\put(136,86){\line(1,0) {25}}
\put(197,122){\line(1,0) {25}}
\put(172,172){\line(1,0) {25}}
\put(233,208){\line(1,0) {25}}
\put(208,283){\line(1,0) {25}}
\put(269,319){\line(1,0) {25}}
\end{picture}

The toric diagram which geometrically engineers $SU(N)^{M-1}$linear quiver gauge theory,
 is constructed from M(i=0,1,2,...,M-1) strip geometries glued together,
 with (i-i)th and i-th strip geometry being the left 
 and right strip, respectively, in the figure 1. The preferred direction
 is the horizontal direction which is indicated by thick lines.
The K\"{a}hler moduli of the homology 
 cycles corresponding to the gluing internal lines of (i-1)th and i-th strip geometries, 
 are $Q^{(i)}_{B_1}, ...,Q^{(i)}_{B_N}$, and the Young tableaux are $Y^{(i)}_1,...,Y^{(i)}_N$. 
 The Young tableau for the left-most horizontal lines, $Y^{(0)}_1,...,Y^{(0)}_N$ and 
 the right-most horizontal lines, $Y^{(M-1)}_1,...,Y^{(M-1)}_N$ are set to be the empty 
 tableaux $\emptyset$.  
 The refined topological string (topological A-model) partition function constructed 
on this toric diagram of figure 1, is found to be,
\[Z^{string}_1=...\sum_{Y^{(i)}_1,...,Y^{(i)}_N}\prod^N_{\alpha=1}(-Q^{(i)}_{B_{\alpha}})^{|Y^{(i)}_{\alpha}|} 
\times H^{Y^{(i-1)}_1,...,Y^{(i-1)}_N}_{Y^{(i)}_1,...,Y^{(i)}_N}(Q^{(i-1)}_{m_1}, Q^{(i-1)}_{F_1},...,Q^{(i-1)}_{m_N})\]
\[ \hspace{1cm} \times H^{Y^{(i)}_1,...,Y^{(i)}_N}_{Y^{(i+1)}_1,...,Y^{(i+1)}_N }(Q^{(i)}_{m_1},Q^{(i)}_{F_1},...,Q^{(i)}_{m_N})... 
\hspace{3cm} (4)\]with\[Y^{(0)}_{\alpha}=\emptyset, Y^{(m-1)}_{\alpha}=\emptyset,  \alpha=1,2,...,N.\]
where the ``local structure" of the whole formula is indicated, which means that only 
factors depending on $Y^{(i)}_{\alpha},(\alpha=1,2,...,N)$are written explicitly whereas the
 other factors are suppressed. \\

The full partition functions should be invariant under
  a change of the choice of the preferred direction, which is called ``slicing invariance conjecture"
  and has been demonstrated to be true in many concrete examples.[3],[4]
  Based on this conjecture, we change the preferred direction from the horizontal
direction to the vertical direction. (Figure 2)\\

Then the partition function is found to be,
 \[Z^{string}_{2}=...\sum_{X^{(0)}_{\alpha},...,X^{(M-1)}_{\alpha}}
\prod^{M-1}_{i=0}(-Q^{(i)}_{F_{\alpha}})^{|X^{(i)}_{\alpha}|} \] \[ \times
H^{X^{(0)}_{\alpha-1},...,X^{(M-1)}_{\alpha-1}}_{X^{(0)}_{\alpha},...,X^{(M-1)}_{\alpha}} 
(Q^{(0)}_{m_{\alpha}},Q^{(1)}_{B_{\alpha}},Q^{(1)}_{m_{\alpha}},Q^{(2)}_{B_{\alpha}},...,Q^{(M-1)}_{B_{\alpha}},Q^{(M-1)}_{m_{\alpha}}) \] 
\[ \times H^{X^{(0)}_{\alpha},...,X^{(M-1)}_{\alpha}}_{X^{(0)}_{\alpha+1},...,X^{(M-1)}_{\alpha+1}}
(Q^{(0)}_{m_{\alpha+1}},Q^{(1)}_{B_{\alpha+1}},Q^{(1)}_{m_{\alpha+1}},Q^{(2)}_{B_{\alpha+1}},...,Q^{(M-1)}_{B_{\alpha+1}}
,Q^{(M-1)}_{m_{\alpha+1}})...\hspace{1cm} (5) \]and the corresponding diagram is depicted in figure 2.\\
\newpage
\begin{picture}(350,320)
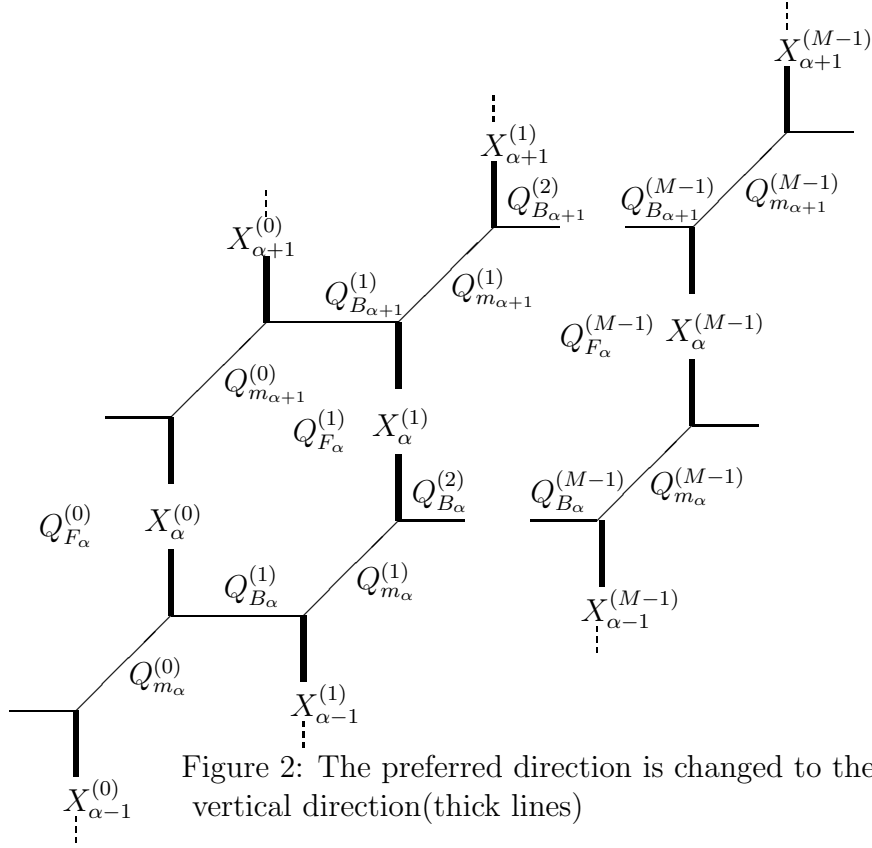
\put(90,25){Figure 2: The preferred direction is changed to the}
\put(90,10){ vertical direction(thick lines)}
\put(50,50){\line(1,1){36}} \put(136,86){\line(1,1){36}} \put(247,122){\line(1,1){36}} \put(86,161){\line(1,1){36}}
\put(172,197){\line(1,1){36}} \put(283,233){\line(1,1){36}}

\put(86,86){\line(1,0){50}} \put(122,197){\line(1,0){50}}

\put(25,50){\line(1,0){25}} \put(172,122){\line(1,0){25}} \put(222,122){\line(1,0){25}} \put(283,158){\line(1,0){25}}
\put(61,161){\line(1,0){25}} \put(208,233){\line(1,0){25}} \put(258,233){\line(1,0){25}} \put(319,269){\line(1,0){25}}

\put(45,12){$X^{(0)}_{\alpha-1}$} \put(50,0){\line(0,1){2}}  \put(50,4){\line(0,1){2}}   \put(50,8){\line(0,1){2}}
\put(131,48){$X^{(1)}_{\alpha-1}$} \put(136,36){\line(0,1){2}}  \put(136,40){\line(0,1){2}}   \put(136,44){\line(0,1){2}}
\put(241,84){$X^{(M-1)}_{\alpha-1}$} \put(247,72){\line(0,1){2}}  \put(247,76){\line(0,1){2}}   \put(247,80){\line(0,1){2}}

\put(107,225){$X^{(0)}_{\alpha+1}$} \put(122,237){\line(0,1){2}} \put(122,241){\line(0,1){2}}  \put(122,245){\line(0,1){2}} 
\put(203,260){$X^{(1)}_{\alpha+1}$} \put(208,273){\line(0,1){2}}  \put(208,277){\line(0,1){2}}   \put(208,281){\line(0,1){2}}
\put(314,296){$X^{(M-1)}_{\alpha+1}$} \put(319,308){\line(0,1){2}}  \put(319,312){\line(0,1){2}}   \put(319,316){\line(0,1){2}}

\put(70,60){$Q^{(0)}_{m_\alpha}$} \put(156,96){$Q^{(1)}_{m_\alpha}$}  \put(267,132){$Q^{(M-1)}_{m_\alpha}$}
\put(106,171){$Q^{(0)}_{m_{\alpha+1}}$} \put(192,207){$Q^{(1)}_{m_{\alpha+1}}$}  \put(303,243){$Q^{(M-1)}_{m_{\alpha+1}}$}
 
\put(106,93){$Q^{(1)}_{B_{\alpha}}$} \put(177,129){$Q^{(2)}_{B_{\alpha}}$}  \put(222,129){$Q^{(M-1)}_{B_{\alpha}}$}
\put(144,204){$Q^{(1)}_{B_{\alpha+1}}$} \put(213,240){$Q^{(2)}_{B_{\alpha+1}}$}  \put(256,240){$Q^{(M-1)}_{B_{\alpha+1}}$}

\put(36,116){$Q^{(0)}_{F_{\alpha}}$} \put(132,152){$Q^{(1)}_{F_{\alpha}}$}  \put(233,188){$Q^{(M-1)}_{F_{\alpha}}$}
\put(76,118){$X^{(0)}_{\alpha}$} \put(162,154){$X^{(1)}_{\alpha}$}  \put(273,190){$X^{(M-1)}_{\alpha}$}

\linethickness{2pt}
\put(50,25){\line(0,1){25}} \put(86,86){\line(0,1){25}} \put(136,61){\line(0,1){25}} \put(172,122){\line(0,1){25}}
\put(249,97){\line(0,1){25}} \put(283,158){\line(0,1){25}} \put(86,136){\line(0,1){25}} \put(122,197){\line(0,1){25}}
\put(172,172){\line(0,1){25}} \put(208,233){\line(0,1){25}} \put(283,208){\line(0,1){25}} \put(319,269){\line(0,1){25}}

\end{picture}

The duality map of the reflection transformation is  given by[1],
\[ (Q^{(\alpha-1)}_{m_i})_d=Q^{(i-1)}_{m_\alpha}, (Q^{(\alpha-1)}_{F_i})_d=Q^{(i)}_{B_{\alpha} },
(Q^{(\alpha)}_{B_i})_d=Q^{(i-1)}_{F_{\alpha}},(Q^{(i)}_{\alpha,\alpha+1})_d=Q^{(\alpha-1)}_{m_{i+1}}Q^{(\alpha)}_{B_{i+1}}\] 
and, \[(X^{(i-1)}_{\alpha})_d=Y^{(\alpha)}_i \]where, the suffix d stands for the dual. 
Then after the duality map, the partition function turns out to be,
\[Z^{string}_3=...\sum_{Y^{(\alpha)}_1,...,Y^{(\alpha)}_M}\prod ^{M-1}_{i=0}(-Q^{(\alpha)}_{B_{i+1}}) \]
\[\times H^{Y^{(\alpha-1)}_1,...,Y^{(\alpha-1)}_M}_{Y^{(\alpha)}_1,...,Y^{(\alpha)}_M}
(Q^{(\alpha-1)}_{m_1}, Q^{(\alpha-1)}_{F_1},..., Q^{(\alpha-1)}_{m_M})\]
\[\times H^{Y^{(\alpha)}_1,...,Y^{(\alpha)}_M}_{Y^{(\alpha+1)}_1,...,Y^{(\alpha+1)}_M} 
(Q^{(\alpha)}_{m_1},Q^{(\alpha)}_{F_1},...,Q^{(\alpha)}_{m_M})... \hspace{3cm} (6) \]
and the corresponding diagram is depicted in figure 3.
 The slicing independence implies that
$Z^{string}_1=Z^{string}_2$ and the duality implies that $Z^{string}_2=Z^{string}_3$.

\begin{picture}(350,320)
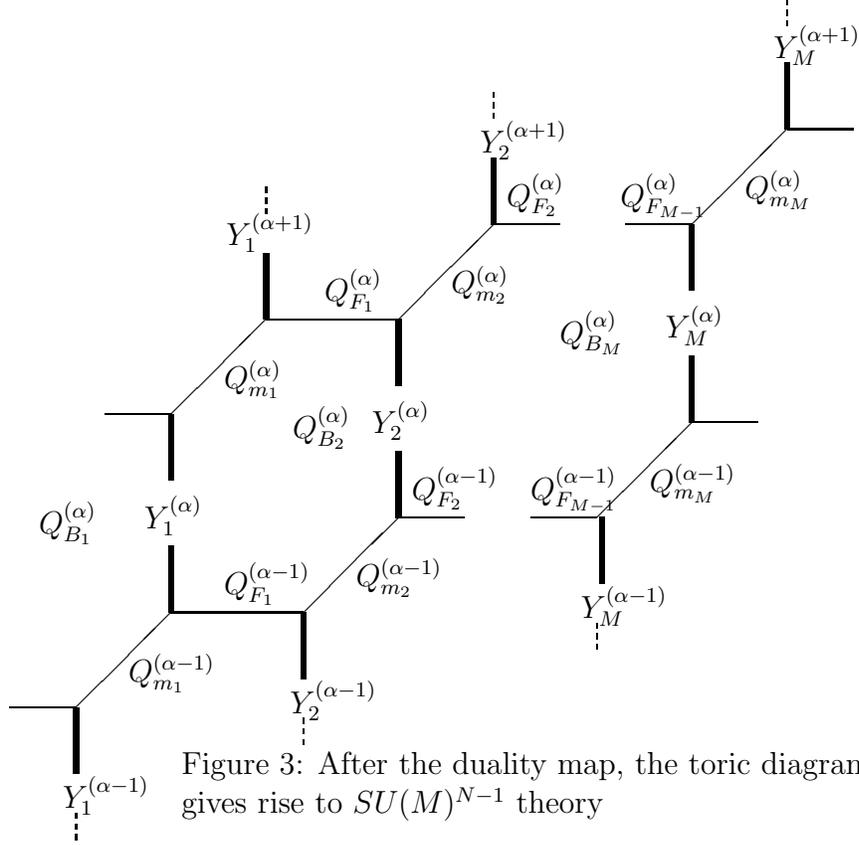
\put(90,25){Figure 3: After the duality map, the toric diagram}
\put(90,10){gives rise to $SU(M)^{N-1}$ theory}
 
\put(50,50){\line(1,1){36}} \put(136,86){\line(1,1){36}} \put(247,122){\line(1,1){36}} \put(86,161){\line(1,1){36}}
\put(172,197){\line(1,1){36}} \put(283,233){\line(1,1){36}}

\put(86,86){\line(1,0){50}} \put(122,197){\line(1,0){50}}

\put(25,50){\line(1,0){25}} \put(172,122){\line(1,0){25}} \put(222,122){\line(1,0){25}} \put(283,158){\line(1,0){25}}
\put(61,161){\line(1,0){25}} \put(208,233){\line(1,0){25}} \put(258,233){\line(1,0){25}} \put(319,269){\line(1,0){25}}

\put(45,12){$Y_{1}^{(\alpha-1)}$} \put(50,0){\line(0,1){2}}  \put(50,4){\line(0,1){2}}   \put(50,8){\line(0,1){2}}
\put(131,48){$Y_{2}^{(\alpha-1)}$} \put(136,36){\line(0,1){2}}  \put(136,40){\line(0,1){2}}   \put(136,44){\line(0,1){2}}
\put(241,84){$Y_{M}^{(\alpha-1)}$} \put(247,72){\line(0,1){2}}  \put(247,76){\line(0,1){2}}   \put(247,80){\line(0,1){2}}

\put(107,225){$Y_{1}^{(\alpha+1)}$} \put(122,237){\line(0,1){2}} \put(122,241){\line(0,1){2}}  \put(122,245){\line(0,1){2}} 
\put(203,260){$Y_{2}^{(\alpha+1)}$} \put(208,273){\line(0,1){2}}  \put(208,277){\line(0,1){2}}   \put(208,281){\line(0,1){2}}
\put(314,296){$Y_{M}^{(\alpha+1)}$} \put(319,308){\line(0,1){2}}  \put(319,312){\line(0,1){2}}   \put(319,316){\line(0,1){2}}

\put(70,60){$Q^{(\alpha-1)}_{m_{1}}$} \put(156,96){$Q^{(\alpha-1)}_{m_{2}}$}  \put(267,132){$Q^{(\alpha-1)}_{m_{M}}$}
\put(106,171){$Q^{(\alpha)}_{m_{1}}$} \put(192,207){$Q^{(\alpha)}_{m_{2}}$}  \put(303,243){$Q^{(\alpha)}_{m_{M}}$}
 
\put(106,93){$Q^{(\alpha-1)}_{F_1}$} \put(177,129){$Q^{(\alpha-1)}_{F_2}$}  \put(222,129){$Q^{(\alpha-1)}_{F_{M-1}}$}
\put(144,204){$Q^{(\alpha)}_{F_1}$} \put(213,240){$Q^{(\alpha)}_{F_{2}}$}  \put(256,240){$Q^{(\alpha)}_{F_{M-1}}$}

\put(36,116){$Q^{(\alpha)}_{B_{1}}$} \put(132,152){$Q^{(\alpha)}_{B_{2}}$}  \put(233,188){$Q^{(\alpha)}_{B_{M}}$}
\put(76,118){$Y^{(\alpha)}_1$} \put(162,154){$Y^{(\alpha)}_2$}  \put(273,190){$Y^{(\alpha)}_M$}

\linethickness{2pt}
\put(50,25){\line(0,1){25}} \put(86,86){\line(0,1){25}} \put(136,61){\line(0,1){25}} \put(172,122){\line(0,1){25}}
\put(249,97){\line(0,1){25}} \put(283,158){\line(0,1){25}} \put(86,136){\line(0,1){25}} \put(122,197){\line(0,1){25}}
\put(172,172){\line(0,1){25}} \put(208,233){\line(0,1){25}} \put(283,208){\line(0,1){25}} \put(319,269){\line(0,1){25}}

\end{picture}\\

 If we set all the $Y$'s 
in $H$'s to the empty tableau $\emptyset$,the perturbative part of the partition function Z is obtained.
Therefore, the normalized amplitude $\tilde{H}$, which is $H$ divided by $H^{\emptyset \emptyset...}_
{\emptyset \emptyset...}$ corresponds to the instanton contribution. 
The normalized amplitudes 
$\tilde{H}$ can be expressed in terms of the "Nekrasov factor" defined by,
\[N_{Y_{\alpha},Y_{\beta}}(t,q,Q)=\prod_{(i,j)\in Y_{\alpha}}
(1-Q q^{Y_{\alpha,i}-j+1}t^{Y^t_{\beta,j}-i})\prod_{(i,j)\in Y_{\beta}}
(1-Q q^{-Y_{\beta,i}+j}t^{-Y^t_{\alpha,j}+i-1}) \]
by virtue of the relation,
\[ \frac{[Y_{\alpha},Y^t_{\beta}]_Q}{[\emptyset ,\emptyset ]_Q}
=\prod_{i,j}\frac{(1-Q q^{i-\frac{1}{2}-Y_{\alpha,j}}t^{j-\frac{1}{2}-Y^t_{\beta,i}})}
{(1-Q q^{i-\frac{1}{2}}t^{j-\frac{1}{2}})}
=N_{Y_{\beta},Y_{\alpha}}(t,q,Q\sqrt{\frac{t}{q}}) \hspace{1cm} (7) \]
Now  we find that the normalized partition function $\tilde{Z}^{string}_1$,  for the diagram in figure 1,
which is $Z^{string}_1$ with all the $H$ factors replaced by $\tilde{H}$ 
is;
\[\tilde{Z}^{string}_1=...\sum_{Y^{(i)}_1,...,Y^{(i)}_N}\prod^{N}_{\alpha=1}(-Q^{(i)}_{B_{\alpha}})^{|Y^{(i)}_{\alpha}|}\]\[\times 
\frac{\prod^N_{\alpha=1} t^{\frac{||Y^{(i)t}_{\alpha}||^2}{2}} \tilde{Z}_{Y^{(i)t}_{\alpha}}(q,t) q^{\frac{||Y^{(i)}_{\alpha}||^2}{2}} 
\tilde{Z}_{Y^{(i)}_{\alpha}}(t,q)}{\prod_{1\leq\alpha<\beta\leq N} N_{{Y^{(i)}_{\beta},Y^{(i)}_{\alpha}}}  (t,q,Q^{(i)}_{\alpha\beta}) 
N_{Y^{(i)}_{\beta},Y^{(i)}_{\alpha}}(t,q,\frac{t}{q}Q^{(i)}_{\alpha\beta})}\]
\[\times \prod_{1\leq\alpha<\beta\leq N} N_{Y^{(i-1)}_{\beta},Y^{(i)}_{\alpha}}(t,q,Q^{(i-1)-1}_{m_{\alpha}}Q^{(i-1)}_
{\alpha\beta}\sqrt{\frac{t}{q}}) N_{Y^{(i)}_{\beta},Y^{(i+1)}_{\alpha}}
(t,q,Q^{(i)-1}_{m_{\alpha}}Q^{(i)}_{\alpha\beta}\sqrt{\frac{t}{q}})\] 
\[\times \prod_{1\leq\alpha\leq\beta\leq N} N_{Y^{(i)}_{\beta},Y^{(i-1)}_{\alpha}}
(t,q,Q^{(i-1)}_{\alpha\beta}Q^{(i-1)}_{m_{\beta}}\sqrt{\frac{t}{q}})
N_{Y^{(i+1)}_{\beta},Y^{(i)}_{\alpha}}(t,q,Q^{(i)}_{\alpha\beta}Q^{(i)}_{m_[\beta}\sqrt{\frac{t}{q}})...\hspace{.1cm} (8)\]
where, the factors which depend on $\vec{Y}^{(i)}$ are written explicitly, whereas those do not, are suppressed.\\

In the gauge theory side, Nekrasov instanton partition function[2] for circle-compactified, 5d $\mathcal{N}=1,
SU(N)^{M-1}$ linear quiver gauge theory is,
\[Z^{SU(N)^{M-1}}_{inst}=...,\sum_{\vec{Y}^{(i)}}(q^{(i)})^{{\sum_{\alpha}}|Y^{(i)}_{\alpha}|}\]
\[\times Z_{bifund}(\vec{a}^{(i-1)},\vec{Y}^{(i-1)},\vec{a}^{(i)},\vec{Y}^{(i)},m^{(i,i-1)},\epsilon_1,\epsilon_2;\beta)\]
\[\times Z_{vect}(\vec{a}^{(i)},\vec{Y}^{(i)},\epsilon_1,\epsilon_2;\beta)\]
\[\times Z_{bifund}(\vec{a}^{(i)},\vec{Y}^{(i)},\vec{a}^{(i+1)},\vec{Y}^{(i+1)},m^{(i,i+1)},\epsilon_1,\epsilon_2;\beta)...\]
where, the bifundamental hypermultiplet contribution is,
\[Z_{bifund}(\vec{a}^{(i-1)},\vec{Y}^{(i-1)},\vec{a}^{(i)},\vec{Y}^{(i)},m^{(i-1,i)},\epsilon_1,\epsilon_2;\beta)\]
\[=\prod^N_{\alpha,\beta=1} P^{-1}_{Y^{(i)}_{\beta},Y^{(i-1)}_{\alpha}}(\epsilon_1,\epsilon_2,
\exp{(-\beta(a^{(i-1)}_{\alpha}-a^{(i)}_{\beta}-m^{(i-1,i)})})\]
For $i=1$ and $i=M$, this reduces to (anti-)fundamental hypermultiplet contribution.
The vector multiplet contribution is,
\[Z_{vect}(\vec{a}^{(i)},\vec{Y}^{(i)},\epsilon_1,\epsilon_2;\beta)
=\prod^N_{\alpha,\beta=1} P^{-1}_{Y^{(i)}_{\beta},Y^{(i)}_{\alpha}}(\epsilon_1,\epsilon_2,
exp(-\beta(a^{(i)}_{\alpha}-a^{(i)}_{\beta}))\]
where, $P^{-1}$ factor is defined by,
\[P^{-1}_{Y_{\alpha},Y_{\beta}}(\epsilon_1,\epsilon_2,a)=\prod_{(i,j)\in Y_{\alpha}} 
sinh\frac{\beta}{2}(a+\epsilon_1(Y_{\alpha,i}-j+1)-\epsilon_2(Y^t_{\beta,j}-i))\]
\[\times \prod_{(i,j)\in Y_{\beta}}sinh\frac{\beta}{2}(a+\epsilon_1(-Y_{\beta,i}+j)
-\epsilon_2(-Y^t_{\alpha,j}+i-1))\hspace{1cm} (9)\]
Plugging in all the bifundamental  hypermultiplet and vector multiplet
contributions, we find that the Nekrasov instanton partition function is,
\[Z^{SU(N)^{M-1}}_{inst}=...\sum_{\vec{Y}^{(i)} }(q^{(i)})^{\sum_{\alpha}|Y^{(i)}_{\alpha}|}
\prod_{1\leq\alpha<\beta\leq N} P_{Y{(i)}_{\beta},Y^{(i)}_{\alpha}}
(\epsilon_1,\epsilon_2,\exp(-\beta(a^{(i)}_{\alpha}-a^{(i)}_{\beta}))\]
\[\times \prod_{1\leq\alpha<\beta\leq N} P_{Y^{(i)}_{\alpha},Y^{(i)}_{\beta}}(\epsilon_1,\epsilon_2,\exp(-\beta
(a^{(i)}_{\beta}-a^{(i)}_{\alpha}))\prod^N_{\alpha=1} P_{Y^{(i)}_{\alpha},Y^{(i)}_{\alpha}}(\epsilon_1,\epsilon_2,1)\] 
\[\times  \prod_{1\leq\alpha<\beta\leq N}(-1)^{|Y^{(i)}_{\alpha}|+|Y^{(i-1)}_{\beta}|}
P^{-1}_{Y^{(i-1)}_{\beta},Y^{(i)}_{\alpha}}(\epsilon_1,\epsilon_2,\frac{t}{q}\exp(-\beta(a^{(i-1)}_{\beta}+
a^{(i)}_{\alpha}+m^{(i-1,i)}))\]
\[\times \prod_{1\leq\alpha<\beta\leq N} (-1)^{|Y^{(i+1)}_{\alpha}|+|Y^{(i)}_{\beta}|}
P^{-1}_{Y^{(i)}_{\beta},Y^{(i+1)}_{\alpha}}(\epsilon_1,\epsilon_2,\frac{t}{q}\exp(-\beta(-a^{(i)}_{\beta}+
a^{(i+1)}_{\alpha}+m^{(i,i+1)}))\]
\[\times \prod_{1\leq\alpha\leq\beta\leq N} P^{-1}_{Y^{(i)}_{\beta},Y^{(i-1)}_{\alpha}}
(\epsilon_1,\epsilon_2,\exp(-\beta(a^{(i-1)}_{\alpha}-a^{(i)}_{\beta}-m^{(i-1,i)}))\]
\[\times \prod_{1\leq\alpha\leq\beta\leq N} P^{-1}_{Y^{(i+1)}_{\beta},Y^{(i)}_{\alpha}}
(\epsilon_1,\epsilon_2,\exp(-\beta(a^{(i)}_{\alpha}-a^{(i+1)}_{\beta}-m^{(i,i+1)}))...\hspace{1cm} (10)\]
Comparing this gauge theory partition function with string partition function $\tilde{Z}^{string}_1$,
we observe that the relations between string moduli and gauge moduli are deformed 
from the self-dual case, and they depend on $\Omega$ background parameters $q$ and $t$.
\[Q^{(i)}_{m_{\alpha}}=\sqrt{\frac{q}{t}}\frac{\tilde{a}^{(i)}_{\alpha}}{\tilde{a}^{(i+1)}_{\alpha}}=
\sqrt{\frac{q}{t}}\exp(-(\beta(a^{(i)}_{\alpha}-a^{(i+1)}_{\alpha}-m^{(i,i+1)}))\]
\[Q^{(i)}_{F_{\alpha}}=\sqrt{\frac{t}{q}}\frac{\tilde{a}^{(i+1)}_{\alpha}}{\tilde{a}^{(i)}_{\alpha+1}}
=\sqrt{\frac{t}{q}}\exp(-\beta(-a^{(i)}_{\alpha+1}+a^{(i+1)}_{\alpha}+m^{(i,i+1)}))\]
where,\[Q^{(i)}_{F_{\alpha}}=Q^{(i)-1}_{m_{\alpha}}Q^{(i)}_{\alpha,\alpha+1}\]
and,\[Q^{(i)}_{\alpha\beta}=\exp(-\beta(a^{(i)}_{\alpha}-a^{(i)}_{\beta})\hspace{2cm} (11)\]
(This does not depend on $q$ or $t$.)\\
If we relate coupling constants $q^{(i)}$ with string theory moduli by,
\[q^{(i)}\prod_{\alpha}\sqrt{\frac{Q^{(i-1)}_{m_{\alpha}}}{Q^{(i)}_{m_\alpha}}}=Q^{(i)}_BQ^{(i-1)}_{m_1}\]
and,\[Q^{(i)}_{B_{\alpha}}=Q^{(i)}_{B_{\alpha-1}}\frac{Q^{(i)}_{m_{\alpha-1}}}{Q^{(i-1)}_{m_{\alpha}}}\]
The relation between $Q^{(i)}_B$ and $q^{(i)}$ is modified from the self-dual case and found to be,
\[Q^{(i)}_B=\sqrt{\frac{t}{q}}q^{(i)}\frac{\tilde{a}^{(i)}_1}{\tilde{a}^{(i-1)}_1}
\prod^N_{\alpha=1}\frac{\sqrt{\tilde{a}^{(i-1)}_{\alpha}\tilde{a}^{(i+1)}_{\alpha}}}{\tilde{a}^{(i)}_{\alpha}}\]
With these identifications, the present author has proved that

 $\tilde{Z}^{string}_1=Z^{SU(N)^{M-1}}_{inst}$
 ant that $\tilde{Z}^{string}_3=Z^{SU(M)^{N-1}}_{inst}$.
 The detail will be reported elsewhere.[6]\\
 
 The duality map of the string theory moduli impiles the following duality map
 of the gauge theory moduli.
 \[(\sqrt{\frac{q}{t}}\frac{\tilde{a}^{(\alpha-1)}_i}{\tilde{a}^{(\alpha)}_i})_d=\sqrt{\frac{q}{t}}
(\frac{\tilde{a}^{(i-1)}_{\alpha}}{\tilde{a}^{(i)}_{\alpha}})\]
\[(\sqrt{\frac{t}{q}}\frac{\tilde{a}^{(i)}_i}{\tilde{a}^{(0)}_{i+1}})_d=\sqrt{\frac{t}{q}}q^{(i)}
\frac{\tilde{a}^{(i-1)}_1}{\tilde{a}^{(i)}_1}\prod^N_{\alpha=1}\frac{\sqrt{\tilde{a}^{(i-1)}
_{\alpha}\tilde{a}^{(i+1)}_{\alpha}}}{\tilde{a}^{(i)}_{\alpha}}\]
\[(\sqrt{\frac{t}{q}}q^{(i)}\frac{\tilde{a}^{(i)}_1}{\tilde{a}^{(i-1)}_1}\prod^N_{\alpha=1}
\frac{\sqrt{\tilde{a}^{(i-1)}_{\alpha}\tilde{a}^{(i+1)}_{\alpha}}}{\tilde{a}^{(i)}_{\alpha}})_d
=\sqrt{\frac{t}{q}}\frac{\tilde{a}^{(1)}_i}{\tilde{a}^{(0)}_{i+1}}\hspace{1cm} (12)\]
If we consider the duality map which preserves the ratio of $\Omega$ background
parameters, $q$ and $t$,such as $q \rightarrow kq, t \rightarrow kt$ for some 
constant $k$, or $q \rightarrow \frac{1}{t} , t \rightarrow \frac{1}{q}$, then the 
duality map of the gauge theory moduli is not changed from the self-dual case.\\

References\\

[1] L. Bao, E. Pomoni, M. Taki and F. Yagi, arXiv :1112.5228[hep-th]

[2] N. A. Nekrasov, Adv. Theor. Math. Phys. 7(2004)831 [hep-th/0206161]

[3] A. Iqbal, C. Koz\c{c}az  and C. Vafa, JHEP 10(2009)069 [hep-th/0701156]

[4] A. Iqbal, C. Koz\c{c}az and K. Shabbir, arXiv:0803.2260[hep-th]

[5] M. Taki, JHEP 083(2008)048 arXiv:0710.1776

[6] Kei Ito, in preparation

\end{document}